\documentclass{emulateapj}

\shorttitle{Composition of CRs from AGNs}
\shortauthors{E.G. Berezhko }

\begin{document}
\title{Composition of cosmic rays accelerated in active galactic nuclei}

\author{E.G.Berezhko\altaffilmark{1}}
\altaffiltext{1}{Yu.G. Shafer Institute of Cosmophysical Research and Aeronomy,
                     31 Lenin Ave., 677980 Yakutsk, Russia}

\email{berezhko@ikfia.ysn.ru}

\begin{abstract} 
The composition of the overall spectrum of cosmic rays (CRs) is studied
under the assumption that ultra high energy CRs above the energy
$10^{17}$~eV are produced at the shock created by 
the expanding cocoons
around active galactic nuclei (AGNs). It is shown that 
the expected CR composition is characterised by two peaks
in the energy dependence of the mean CR atomic number $<A(\epsilon)>$.
The first one at the energy $\epsilon\approx10^{17}$~eV corresponds 
to the very end of the Galactic CR component, 
produced in supernova remnants (SNRs). It is followed
by a sharp decrease of $<A(\epsilon)>$ within the energy interval
from $10^{17}$ to $10^{18}$~eV. This is a 
signature of the transition from Galactic to
extragalactic CRs. The second peak, with $<\ln A>\approx 2$, at energy 
$\epsilon\approx10^{19}$~eV, expected at the beginning of the GZK cutoff,
is the signature of the CR
production by the nonrelativistic  cocoon shocks.
The calculated CR composition is consistent with the
existing data. The alternative scenario, which suggests reacceleration increasing
the energy of CRs produced in SNRs by a factor of 30, is also examined.
\end{abstract}


%

\section{Introduction}
The overall origin of cosmic rays (CR), in particular at ultra high energies,
above $10^{18}$~eV, is still an unresolved problem in astrophysics
\citep[e.g.][]{NW00,stanev04,cronin05,inoue08}.  Understanding this origin
requires the determination of the astrophysical objects, that are the CR
sources, and of the appropriate acceleration processes, that form the CR
spectrum in these objects. During the last several
years considerable progress has been achieved in this field, both
experimentally and theoretically.  Recently the sharp steepening of the CR
spectrum above $3\times 10^{19}$~eV was established in the HiRes
\citep{hires07} and Auger \citep{augerJ07} experiments.  It presumably
corresponds to the so-called Greizen-Zatsepin-Kuzmin (GZK) cutoff, caused by
energy losses of the CRs in their interactions with the microwave
background radiation.  This is evidence that the highest energy part of the CR
spectrum is of extragalactic origin.  It was also recently demonstrated
\citep{bv07} that the CRs with energies up to $\epsilon \sim 10^{17}$~eV can be
produced in supernova remnants (SNRs) as a result of diffusive shock
acceleration \citep{krym77,bell78}, and that the observed CR energy spectrum
can be well represented by two components: the first, dominant up to
$10^{17}$~eV, consists of CRs produced in Galactic SNRs; the second, dominant
at energies above $10^{18}$~eV, is of extragalactic origin.  The latter 
component can be produced at the shock that is formed by the expanding
cocoon around active galactic nuclei (AGNs): above the energy $10^{18}$~eV the
overall energy spectrum of CRs, produced during the AGN evolution and released
into intergalactic space, has an appropriate power law form, which extends at
least up to the energy $\epsilon_\mathrm{max}\sim 10^{20}$~eV, 
well above the GZK cutoff \citep{ber08}. This is the 
``dip scenario'' of the overall CR spectrum formation \citep{berez06}. Within
the alternative ``ankle scenario'', extragalactic CRs are expected to dominate
only above the energy $10^{19}$~eV \citep{berez06}. It was noted
\citep{bv07,ber08}, that the composition of CRs at $\epsilon>10^{17}$~eV is
expected to be very different in these two scenarios.  Therefore the study of
CR composition can in principle provide the signature of the acceleration
mechanism which creates the spectrum of extragalactic CR component.

Here we present the calculations of the expected composition of CRs,
accelerated by cocoon shocks and demonstrate that it has well pronounced
peculiarities, which can be considered as a signature of ultra high energy CR
production by nonrelativistic shocks.  The calculated composition is compared
with the composition expected within the alternative scenario and with the
existing data. The alternative scenario, which suggests reacceleration increasing
the energy of CRs produced in SNRs by a factor of 30, is also examined.

\section{Results and discussion}
The composition of CRs is determined if we know not only their all-particle
spectrum,
\begin{equation}
J(\epsilon)=\sum_A J_\mathrm{A}(\epsilon),
\end{equation}
that is the differential intensity $J(\epsilon)$ with respect to particle
energy $\epsilon$, but also the spectra $J_\mathrm{A}(\epsilon)$ of all relevant
elements with atomic number $A$. Since the shock-accelerated CRs 
come originally from the thermal background plasma, one has
to expect that the flux of each CR element in/near the source, is proportional
to the number density of this element $N_\mathrm{A}$ in the background plasma:
$J_\mathrm{A}^\mathrm{s}(\epsilon)\propto N_\mathrm{A}$.  
Therefore it is useful to represent the spectrum
of each CR element in the form
\begin{equation}
J_\mathrm{A}^\mathrm{s}(\epsilon)=e_\mathrm{A}(\epsilon)
 a_\mathrm{A} J_\mathrm{H}^\mathrm{s}(\epsilon),
\end{equation}
where $a_\mathrm{A}=N_\mathrm{A}/N_\mathrm{H}$ 
is the abundance of element $A$ relative to the hydrogen
abundance and $e_\mathrm{A}(\epsilon)$ is an enrichment factor which describes the
preferential production of element $A$ relative to the production of protons
(marked by the subscript H).

Direct measurements of CR fluxes at energies $\epsilon <10^{14}$~eV show that
the factor $e_\mathrm{A}>1$ considerably exceeds unity and that it progressively
increases with the increase of $A$ \citep[e.g.][]{bk99}. This means a
considerable enrichment of CRs in heavy elements compared with the interstellar
abundance.  Note that at these relatively low energies, where direct
determination of CR particle parameters is possible, the fluxes $J_\mathrm{A}$ 
and $J_\mathrm{H}$
are usually compared at the same energy per nucleon, that is for
$\epsilon_\mathrm{A}=A\epsilon_\mathrm{H}$.  
Since it is impossible to determine the mass of
individual CR particles at ultra high energies, the particle energy instead of
the energy per nucleon is the most appropriate variable for the analysis of CR
spectrum and composition in this case.

The observed CR composition at $\epsilon<10^{17}$~eV
is well consistent with the scenario
in which Galactic CR component is produced by supernova shocks \citep{bv07}.
The enrichment of shock-accelerated CRs by heavy elements is due to a
more effective injection and subsequent acceleration 
of heavy ions \citep[see][for the details]{bk99}. 

The spectra of 
ultrahigh energy CRs produced by nonrelativistic cocoon shocks
can be represented in the form \citep{ber08}
\begin{equation}
J_\mathrm{A}^\mathrm{s}(\epsilon)=C_\mathrm{A}\epsilon^{-\gamma},
\end{equation}
where $\gamma\approx 2.6$ and the values of parameters 
$C_\mathrm{A}\propto e_\mathrm{A}N_\mathrm{A}$ are
determined by the background medium's composition 
$N_\mathrm{A}$ and by the enrichment
factors $e_\mathrm{A}$. 
Since the injection/acceleration of CRs at cocoon and supernova
shocks are expected to be very similar, the values of the enrichment factors
$e_\mathrm{A}$ in the former case can be extracted from the experimentally measured
fluxes of the Galactic CR component 
$J_\mathrm{A}(\epsilon)$ at some fixed energy, say
$\epsilon=1$~TeV \citep[see e.g.][]{bv07}, according to the expression
\begin{equation}
e_\mathrm{A}=
J_\mathrm{A}(1\mbox{~TeV})/J_\mathrm{H}(1\mbox{~TeV})
(N_\mathrm{H}/N_\mathrm{A})_{\odot}.
\end{equation}
Here $(N_\mathrm{A}/N_\mathrm{H})_{\odot}$ is the solar system relative
abundance of elements with atomic number $A$.

In order to calculate the spectra of extragalactic CRs inside the Galaxy
one needs to take into account the modification of the
CR spectra due to their interaction with the cosmic microwave
background (CMB) and due to their modulation in the Galactic wind.
This can be done by representing the observed 
extragalactic CR spectrum in the form
\begin{equation}
J_\mathrm{A}(\epsilon)=
P_\mathrm{l}P_\mathrm{h}J_\mathrm{A}^\mathrm{s}(\epsilon),
\end{equation}
where factors $P_\mathrm{l}$ and 
$P_\mathrm{h}$ describe CR flux modification due to Galactic
wind and CMB respectively. Since the diffusive mobility of the CRs is inversely
proportional to the particle charge number $Z$, the first factor can be
represented in the form
\begin{equation}
P_\mathrm{l}=\exp (-Z\epsilon_\mathrm{l}/\epsilon),
\end{equation}
where $\epsilon_\mathrm{l}\sim 10^{17}$~eV \citep{bv07}.  
This expression shows that at
a given energy the lighter CR species can more easily penetrate the inner
Galaxy from outside. Due to this factor the observed extragalactic CR
component becomes progressively heavier with increasing energy.

To describe the effects of CRs interacting with the CMB radiation during their
propagation from extragalactic sources to the observer 
-- pion and pair production, CR nuclei photodesintegration and adiabatic cooling --
we use a simple
analytical form of the factor $P_\mathrm{h}$
\begin{equation}
P_\mathrm{h}=\exp [-(A\epsilon/\epsilon_\mathrm{h})^2].
\end{equation}
At $\epsilon_\mathrm{h}=1.3\times 10^{18}$~eV 
for the case of helium and iron nuclei it
satisfactorily fits the results of detailed calculations 
which consistently describes all these effects  
\citep{berez06}. In the case of protons, instead of
the modification factor $P_\mathrm{h}$ we directly use their 
modified spectrum calculated by \citet{berez06}. 

%
\begin{figure}[t]
\plotone{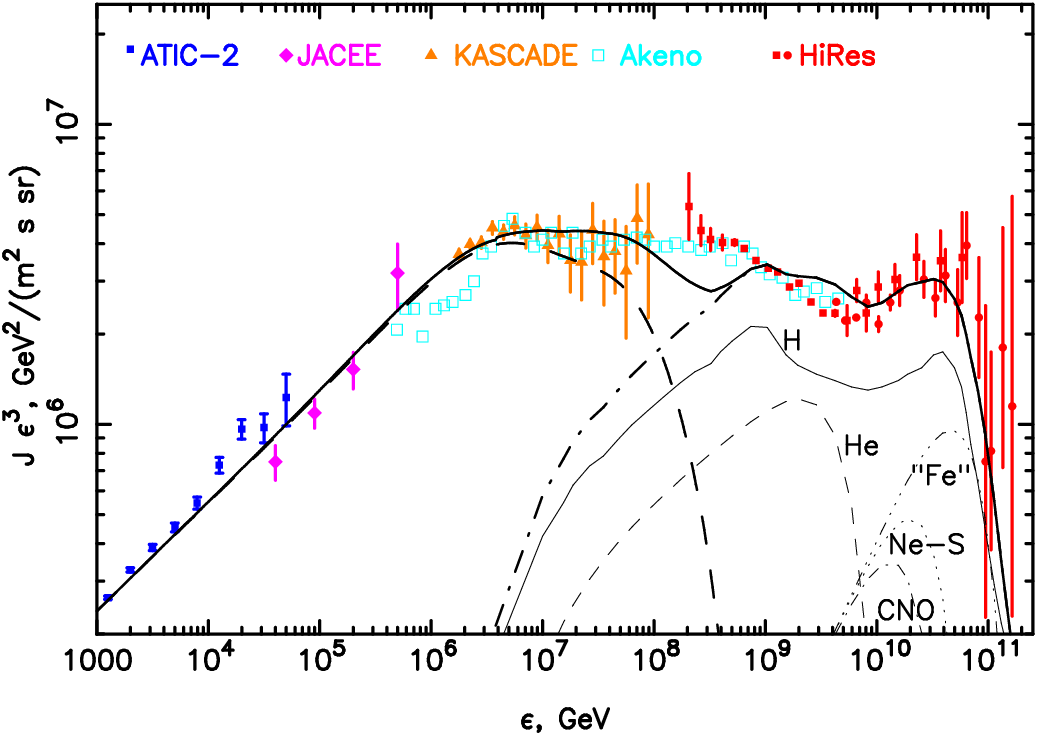} 
\figcaption{Overall CR intensity as a function of
    energy (thick solid line), Galactic CR component produced in SNRs (thick
    dashed line) and extragalactic component produced in the IGM (thick
    dash-dotted line).  The spectra of different extragalactic components are
    shown by thin lines.  Experimental data obtained in the ATIC-2
    \citep{atic2}, JACEE\citep{jacee}, KASCADE \citep{kascade}, Akeno
    \citep{takeda03}, and HiRes \citep{hires07} experiments are shown as well.}
\label{f1}
\end{figure}

As is clear from expression (7), the interaction of CR nuclei with the
CMB radiation produces a sharp cutoff in their spectra at the energy,
which is proportional to the nuclear mass.
Due to this fact CRs are expected to become progressively heavier at energies
above $10^{18}$~eV.

The cocoon shock propagates initially through the interstellar medium (ISM) of
the host galaxy and later through the intergalactic medium (IGM). The IGM is
less abundant in heavy elements than the ISM: the median metallicity of the IGM
is $(N_\mathrm{A}/N_\mathrm{H})=
0.2(N_\mathrm{A}/N_\mathrm{H})_{\odot}$ for all relevant elements heavier than
helium \citep[e.g.][]{cen06}.  Since the maximal CR energy produced by the
expanding cocoon shock decreases with time \citep{ber08}, the local metallicity
at the shock front changes from the ISM value 
$(N_\mathrm{A}/N_\mathrm{H})\approx
(N_\mathrm{A}/N_\mathrm{H})_{\odot}$ 
at early phases, when the shock produces CRs up to
$\epsilon_\mathrm{max} \sim Z\times10^{20}$~eV, to the IGM value at late phases, when
$\epsilon_\mathrm{max} \sim Z\times10^{18}$~eV.  To describe this effect we uses a
metallicity $(N_\mathrm{A}/N_\mathrm{H})=
0.2[\epsilon/(Z\times10^{18}~\mbox{eV})]^{0.35}(N_\mathrm{A}/N_\mathrm{H})_{\odot}$.
For helium we use the cosmological value 
$N_\mathrm{He}=0.08N_\mathrm{H}$ \citep[e.g.][]{berez06}. 
Using these number
densities one can calculate the relative values of the coefficients 
$C_\mathrm{A}/C_\mathrm{H}$
for all elements except protons.  The proton coefficient 
$C_\mathrm{H}$ depends on a
number of physical factors such as the IGM hydrogen number density 
$N_\mathrm{H}$ and the
power and spatial distribution of nearby AGNs. Its value is determined by the
fit of the expected all-particle CR flux $J(\epsilon)$ to the existing
measurements of the CR intensity at energies $\epsilon >10^{18}$~eV.

Secondary protons appearing as a result of photodesintegration of helium
nuclei are also included in an approximate way: it is assumed that the secondary
protons (whose total number equals four times the number of helium nuclei in
the source spectrum with energy above $2\times 10^{18}$~eV) have the normal
energy distribution around $\epsilon_0 =5\times 10^{17}$~eV within the interval
$\Delta\epsilon=\epsilon_0$.

In Fig.1 we present the all-particle spectrum which includes two components:
CRs produced in SNRs \citep{bv07} and extragalactic CRs, which consist of
protons (H), helium (He) and three groups of heavier nuclei, produced in AGNs.
Note, that the dip at energy $\epsilon\approx 10^{19}$~eV is still clearly seen
even though it is less pronounced compared with the case of pure proton
composition.  The calculated overall CR spectrum is in a satisfactory way
consistent with the experimental data, except in  the energy interval
$10^{17}<\epsilon< 10^{18}$~eV in the transition region, where, contrary to the
experiment, the calculated spectrum has a small dip.  
Taking into account the
existing experimental uncertainties at these energies it is not clear
whether this discrepancy is a serious problem for the model or not.  Note, that
the size and the shape of this peculiarity is very sensitive to the details of
the galactic CR high energy cutoff and to the extragalactic low energy spectrum
part. Therefore it is hard to predict it quantitatively.  At the same time the
overall spectrum, which is a mixture of two independent components, should have
such peculiarity within the transition region.  If the actual CR spectrum is
indeed so uniform within the energy range $10^{17}<\epsilon<10^{18}$~eV, as it
looks according to the Akeno data, then it should be considered as a serious
problem for the dip scenario.
 
%
\begin{figure}[t]
\plotone{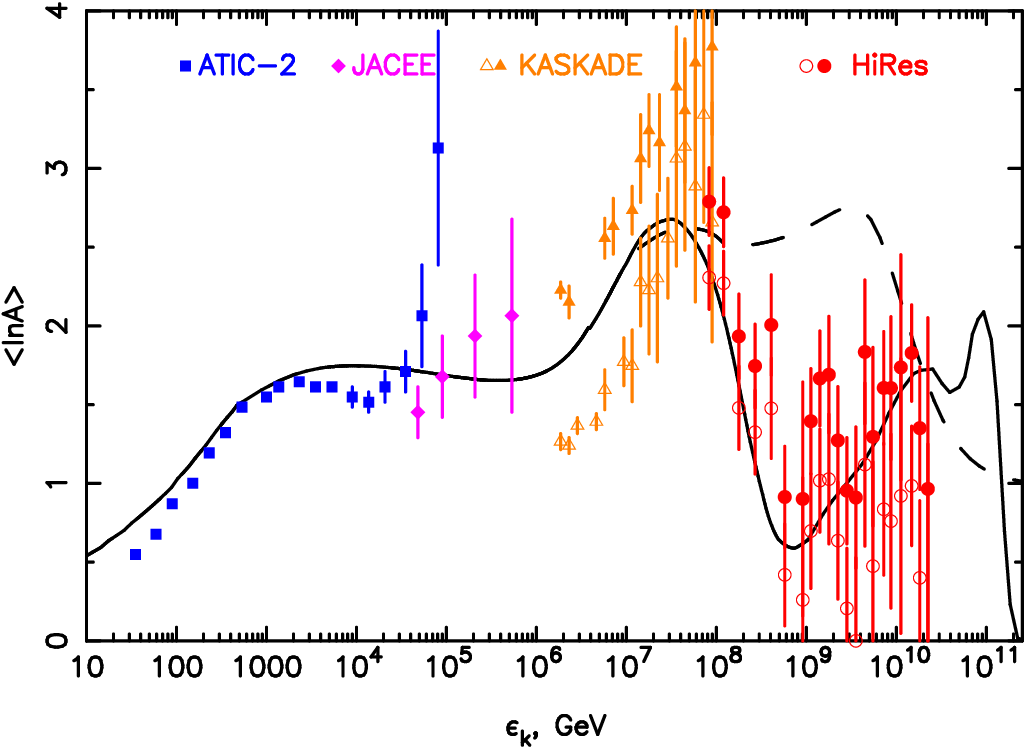}
\figcaption{Mean logarithm of the CR nucleus atomic number as a 
function of energy. Solid and dashed lines correspond to
the dip and ankle scenarios respectively.
Experimental data 
obtained in the ATIC-2, JACEE, KASCADE 
\citep[QGSJET and SYBYLL;][]{hoerandel05}, and
HiRes \citep[QGSJET and SYBYLL;][]{hires05}
experiments are
shown.}
\label{f2}
\end{figure}

The mean logarithm of CR atomic number is represented in Fig.2 as a function of
energy.  As already qualitatively predicted \citep{ber08}, the CR mass energy
dependence $<\ln A(\epsilon)>$ has two peaks.  The first one at the energy
$\epsilon\approx10^{17}$~eV corresponds to the very end of the galactic CR
component \citep{bv07}, whereas the second, at the energy
$\epsilon\approx10^{19}$~eV, is at the beginning of the GZK cutoff.  Note that
at high energies $\epsilon>10^{15}$~eV information about CR composition is
obtained from the mean values of the shower maxima $X_\mathrm{max}$
(usually measured in g/cm$^2$), determined
by the ground based detectors. Knowing the average depth of the shower
maximum for protons, $X_\mathrm{max}^\mathrm{H}$, and for iron nuclei,
$X_\mathrm{max}^\mathrm{Fe}$, from simulations, the mean logarithmic mass can be
estimated from the measured $X_\mathrm{max}$ according to the relation
\citep[e.g.][]{hoerandel05}
\begin{equation}
<\ln A>=(X_\mathrm{max} - X_\mathrm{max}^\mathrm{H})/
(X_\mathrm{max}^\mathrm{Fe} - X_\mathrm{max}^\mathrm{H})<\ln 56>.
\end{equation}
This conversion requires the choice of a particle interaction model. Here we
use the values of $X_\mathrm{max}$, measured in the HiRes experiment
\citep{hires05}, and the QGSJET and SYBYLL models to determine $<\ln A>$.  It
is seen from Fig.2 that the HiRes data are very well consistent with the
expected sharp decrease of $<\ln A>$ within the energy interval
$10^{17}-10^{18}$~eV.  At higher energies, between $10^{18}$ and $10^{19}$~eV,
the experimental values of $<\ln A(\epsilon)>$ have a quite irregular behavior.
Nevertheless, the experiment reveals a trend of progressive increase of the
mean CR mass so that the expected peak value $<\ln A>\approx 1.7$ at the energy
$\epsilon\sim 10^{19}$~eV is consistent with the existing HiRes data. It is
also clear from Fig.2, that a more precise experimental determination of the CR
mass composition for $\epsilon>10^{15}$~eV is required for a more complete
conclusion about the CR origin.

\section{Alternative scenario}
The alternative scenario for the overall CR spectrum is the
``ankle scenario'' \citep{berez06}. In this case the extragalactic source
spectrum, as compared with the dip-scenario, is assumed to be much harder
$J^\mathrm{s}_\mathrm{A}\propto \epsilon^{-2}$ so 
that it becomes dominant above an
energy of $\epsilon=10^{19}$~eV \citep[e.g.][]{berez06}. Therefore, to fit the
observed overall CR spectrum one needs a third component to fill the gap
between the Galactic CR spectrum, produced by SNRs, and the hard extragalactic
spectrum.  It is hard to believe that this third CR component could be
completely independent of the CR component produced in SNRs, since in such a
case one should expect some peculiarity in the spectral shape at energy
$\epsilon\approx 10^{17}$~eV, where these two components are expected to match.
The appropriate solution of this problem is the existence of some kind of
reacceleration process which picks up the most energetic CRs from SNRs and
substantially increases their energy, resulting in a smooth extension of
the first CR component towards the higher energies.

We model the spectra of reaccelerated CRs in the following way,
without specifying the reacceleration mechanism. 
For every element with the nuclear
charge number $Z$, as in the dip scenario, we use the spectrum,
which coincides with
$J_\mathrm{A}(\epsilon)$ for
$\epsilon<\epsilon_\mathrm{max1}^\mathrm{Z}$ and has a form
\begin{equation}
J_\mathrm{A}(\epsilon)=
J_\mathrm{A}(\epsilon_\mathrm{max1}^\mathrm{Z})
(\epsilon/\epsilon_\mathrm{max1}^\mathrm{Z})^{-\gamma}
\exp(-\epsilon/\epsilon_\mathrm{max2}^\mathrm{Z})
\end{equation}
at $\epsilon>\epsilon_\mathrm{max1}^\mathrm{Z}$.  Here
$\epsilon_\mathrm{max1}^\mathrm{Z}$ is the minimum energy of particles involved
in reacceleration, and $\epsilon_\mathrm{max2}^\mathrm{Z}$ is the maximum
particle energy achieved during reacceleration. It is natural to assume that
these energies scale proportional to the rigidity
$\epsilon_\mathrm{max}^\mathrm{Z}=Z\epsilon_\mathrm{max}^\mathrm{p}$. Here the
superscript $p$ denotes protons. The quantities
$\epsilon_\mathrm{max1}^\mathrm{p}$, $\epsilon_\mathrm{max2}^\mathrm{p}$ and
$\gamma$ are treated as free parameters whose values are determined as a result
of the best fit. CR acceleration by spiral density shocks in the Galactic
Wind \citep{vz04}, or acceleration in the vicinity of pulsars
\citep{bell92,ber94} could play the role of the reacceleration mechanism.

\begin{figure}[t] 
\plotone{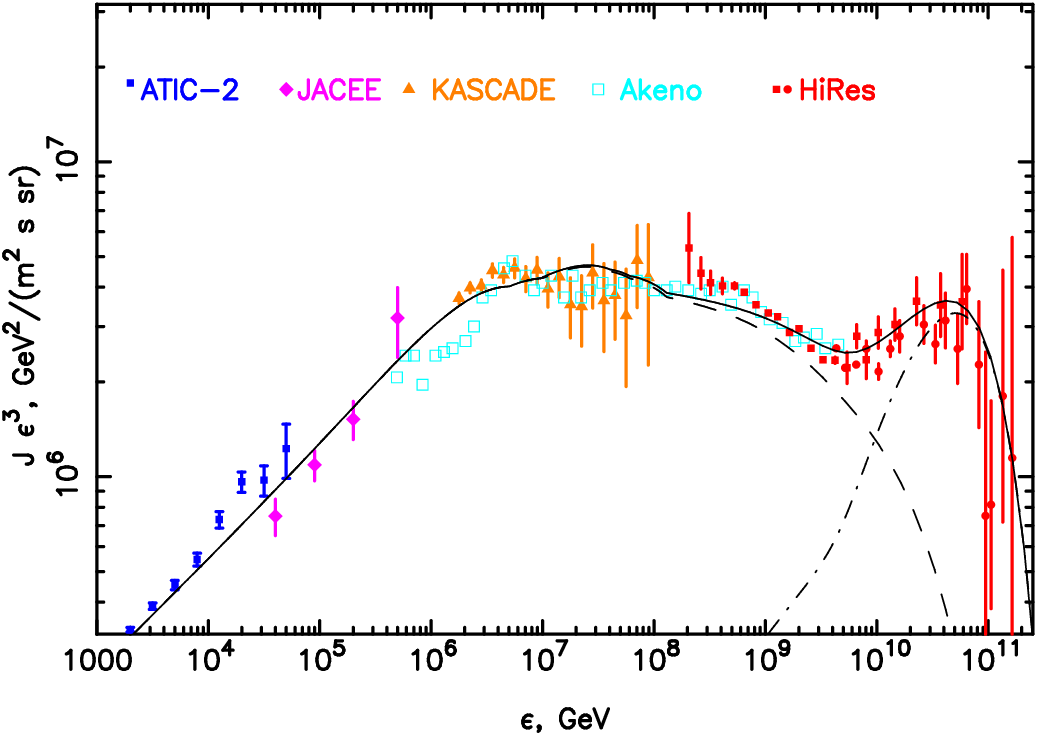} 
\figcaption{The same as in Fig.1, but for the ankle scenario.
The dashed line represents the Galactic component, which includes
CRs produced in SNRs and reaccelerated CRs. The dash-dotted line
represents the extragalactic component; it corresponds to the source
spectrum $J^\mathrm{s}_\mathrm{A}\propto \epsilon^{-2}$ \citep{berez06}.  
}
\label{f3}
\end{figure}

We present in Fig.3 the CR spectrum calculated within the ankle scenario, with
the adopted values $\epsilon_\mathrm{max1}^\mathrm{p}=5\times 10^{15}$~eV,
$\epsilon_\mathrm{max2}^\mathrm{p}=1.5\times 10^{17}$~eV and $\gamma = 3$,
which provide the best fit to the observed CR spectrum. 
Following \citet{berez06} we use the extragalactic CR source spectrum
in the form $J^\mathrm{s}_\mathrm{A}\propto \epsilon^{-2}$ and
assume the composition of these CRs with $<A(\epsilon)>=1$.

It is seen that at $\epsilon >10^{17}$~eV it is well consistent with the data. 
However, such a well-known
peculiarity in CR spectrum as the knee at $\epsilon \approx
3\times 10^{15}$~eV is much less pronounced in the theoretical CR
spectrum than in the experiment. This may be considered as an indication
against the ankle scenario.

As it is seen in Fig.2, the CR composition corresponding to the ankle scenario
is considerably heavier at energies $10^{17}\epsilon< 10^{19}$~eV than in the
dip scenario and  is inconsistent with the HiRes
measurements.  Note, that CR composition of the reaccelerated CRs is a direct
consequence of the heavy CR composition at the energy $\epsilon >\epsilon_\mathrm{max1}^\mathrm{Z}$,
where the reacceleration is assumed to start, and it is not sensitive to the
details of reacceleration process.

Since the expected spectrum of CRs produced in gamma-ray bursts (GRBs) 
has a form $J^\mathrm{s}\propto \epsilon^{-\gamma}$ with
$\gamma= 2-2.2$ only ankle scenario is possible if GRBs are the
the source of extragalactic CR component 
\citep[see][and references therein]{berez06}. It means that HiRes
CR composition is against GRBs as a source of ultra high energy CRs.
In addition, as \citet{berez06} argued, 
the energy output of GRBs has a
serious problem if they are considered as 
the main source of the extragalactic CRs.
Note also, that based on the data collected at the Auger experiment,
correlation between the arrival directions of CRs with energy above $6\times
10^{19}$~eV and the position of nearby AGNs has been find \citep{augercol07},
that strongly supports AGNs as a prime candidate for the source of ultra high
energy CRs.

\section{Summary}

The composition of ultra high energy CRs produced by nonrelativistic cocoon
shocks around AGNs is characterised by well pronounced peculiarities which are
two peaks in the energy dependence of the mean CR atomic number
$<A(\epsilon)>$.  The first peak at the energy $\epsilon\approx10^{17}$~eV
corresponds to the very end of the Galactic CR component, produced in SNRs
\citep{bv07}, whereas the second, at the energy $\epsilon\approx10^{19}$~eV, is
expected at the beginning of the GZK cutoff.  The strong energy dependence of
the CR composition within the energy interval from $10^{17}$ to $10^{18}$~eV is
expected as a signature of the transition from heavy Galactic to light
extragalactic CRs, whereas the detection of a heavy CR composition at
$\epsilon\approx10^{19}$~eV has to be considered as the signature of CR
production by the nonrelativistic cocoon shocks.

The existing measurements of CR composition are consistent with
the dip scenario with a formation of the CR spectrum in SNRs and AGNs and
are inconsistent with the ankle scenario, which includes
reacceleration of CR produced in SNRs as a third CR component. 
Additional peculiarity in the overall CR spectrum -- the dip 
in the transition region $10^{17}<\epsilon<10^{18}$~eV --
is expected within the dip scenario. Since it is not seen in the existing
data it is a real difficulty for the dip scenario.
It is therefore
clear that a more precise measurements of CR spectrum and
composition at energies above
$10^{17}$~eV are needed for a strict determination of CR origin.

\acknowledgements
I thank H.J. V\"olk for the helpful discussions.
The work was supported by the 
Russian Foundation for Basic Research (grants 06-02-96008, 07-02-00221)

\end{document}